# Privilege and confidentiality in generative AI workflows

Václav Janeček, *University of Bristol Law School*

Thomas Melham, *University of Oxford, Department of Computer Science*

**Acknowledgement.** This article arose out of Dr Janeček's research at the IGSG-CNR in Florence (Leverhulme International Fellowship) and the authors' interdisciplinary teaching for the Oxford LawTech Education Programme.

**Abstract.** Generative AI (GenAI) systems store and process client data in three ways: in the model's parameters through training and memorisation, in the context window during a live session, and in knowledge databases for retrieval-augmented generation (RAG). This article explains the three modes in terms accessible to practitioners. Each creates different and often counter-intuitive risks to confidentiality and legal professional privilege, and each calls for a specific governance response. Drawing on the first UK and American decisions to address privilege and generative AI, on the orthodox authorities against which they must be read, and on recent computer science research, we then situate the analysis within the regulatory framework governing solicitors in England and Wales and within the ordinary principles of professional negligence. We show how the standard of effective information governance (and with it the benchmark for negligence and misconduct) is changing. Although we write primarily for SRA-regulated practitioners, the analysis extends to any jurisdiction in which privilege or professional secrecy depends on demonstrable confidentiality.

## 1 Introduction

In generative AI systems, data moves in ways that are often invisible by design. And yet, wherever client data goes, the lawyer's professional obligations go with it. In this article, we examine three distinct ways in which data—including confidential client data and personal data—is stored and processed in generative AI systems: (1) model memorisation; (2) the context window; and (3) retrieval-augmented generation (RAG). For each we identify risks that are easy to miss even for technically informed users, and what practitioners can do about them. Related data protection issues are also discussed.

The headline message is threefold. First, the risks are manageable if one understands the difference between the three modes of data storage and processing (Figure 1 gives an overview of the three modes and their persistence characteristics). A lawyer who cannot distinguish between data that is 'baked into' a model, data that transits a provider's infrastructure during a session, and data that persists in a vector database, cannot make sound judgements about confidentiality, privilege or data protection compliance. That is not a demand that every practitioner audit the technology stack personally. Responsibility is often distributed across procurement, risk and IT. What cannot be delegated, though, is the judgement at the point of AI use. Second, the law of privilege does not need to change to meet these risks, but it does need to be applied accurately. The first English decision to address the point, *Munir* (Section 3.2), treats an upload to a consumer-grade tool as placing the material in the public domain. It does not. Material disclosed to an identified provider on terms permitting retention, training use and onward disclosure loses protection because disclosure on those terms is inconsistent with maintaining confidence as against the provider, not because the material has been published. On the limited-disclosure authorities, the same reasoning places a properly configured enterprise-grade AI deployment on the safe side of the line. Third, because generative AI systems pose new and unintuitive data leakage risks, information governance protocols need to be adjusted accordingly. The standard of what counts as effective data governance and, by extension, what a reasonably competent practitioner must do to protect privileged and confidential material is changing.

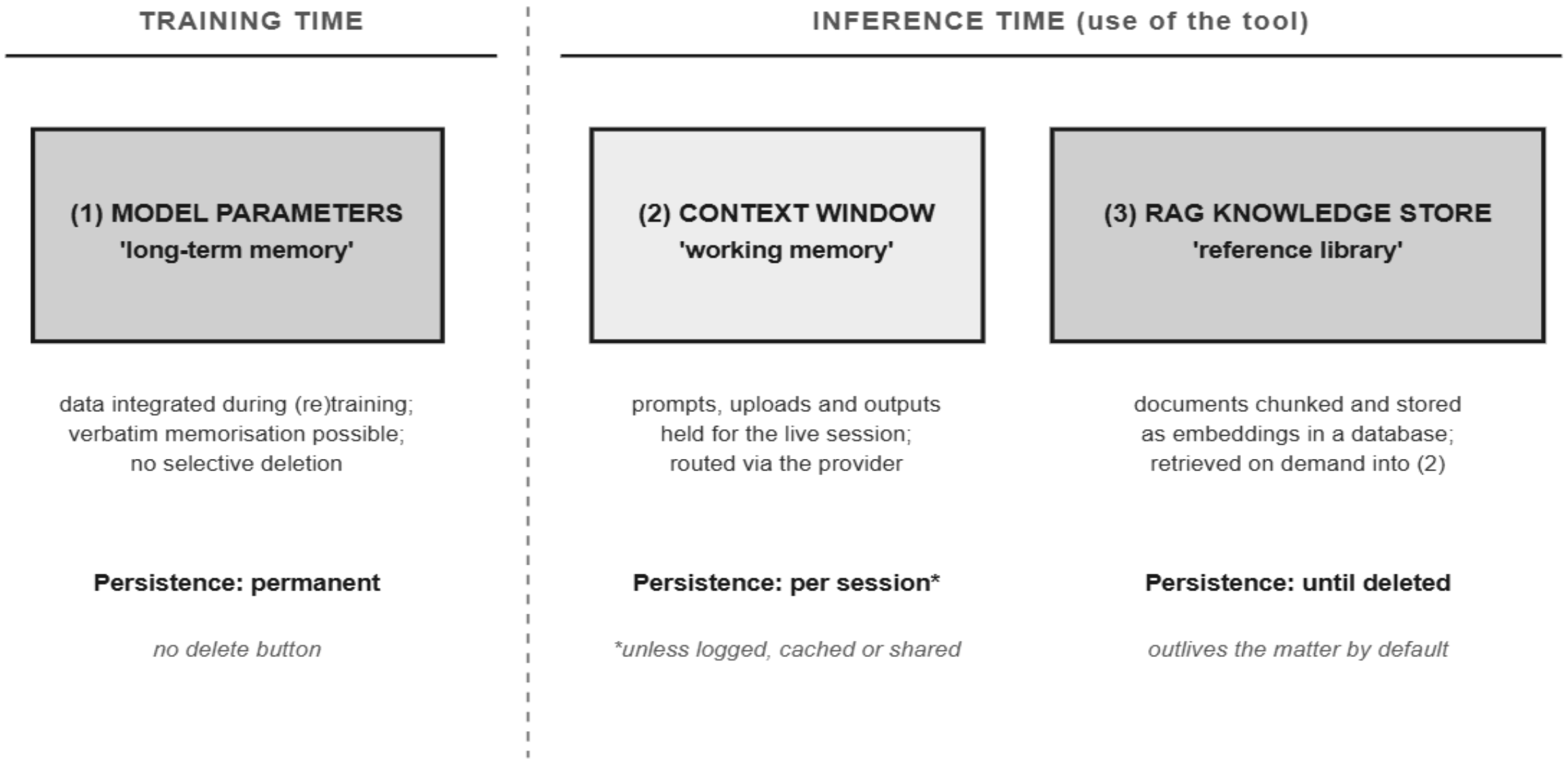

*Figure 1: Three modes of data persistence in generative AI systems*

The legal context has sharpened during the past year. In November 2025 the Upper Tribunal warned in *Munir* that uploading confidential client documents into a consumer-grade AI tool 'is to place this information on the internet in the public domain, and thus to breach client confidentiality and waive legal privilege'.[1] That warning was elevated into the formal headnote and later endorsed by the SRA's warning notice,[2] though it was strictly obiter, as discussed below. In February 2026, in *United States v Heppner*, the Southern District of New York held—as 'a question of first impression nationwide'—that a defendant's exchanges with a consumer-grade chatbot were protected neither by attorney–client privilege nor by the work product doctrine.[3] In April 2026, Sir Colin Birss (Chancellor of the High Court, as he then was) stressed in a speech that AI does not change the legal test for privilege but that confidentiality—its precondition—cannot be assumed when consumer-grade systems are used.[4]

[1] *UK and R (on the application of Munir) v Secretary of State for the Home Department (AI hallucinations; supervision; Hamid)* [2026] UKUT 81 (IAC) at [60] (hereafter *Munir*).

[2] SRA 'Misuse of AI' (warning notice, 17 August 2026), available at https://perma.cc/BTJ3-VTZB.

The doctrinal framework is familiar, but the technology to which it must now be applied is not. Our purpose is to make that technology visible. Some of these issues have been analysed in emerging practitioner and academic writings,[5] but this article goes further, addressing the technical understanding and AI literacy which, if lacking, is likely to become the source of professional negligence in this field.

Some terminological clarifications are needed at the outset. First, loose labels such as 'open source' or 'closed source' AI tools have already caused judicial confusion (as will be seen) and we thus distinguish between *consumer-grade* tools—services used on standard public terms, under which the provider may typically retain inputs, use them for model training (subject to opt-outs), and disclose them in defined circumstances—and *enterprise-grade* deployments, in which inputs are processed under a negotiated contractual framework that typically excludes training use, restricts retention, and imposes confidentiality obligations on the provider. 'Enterprise' can be a commercial label rather than a guarantee, but the distinction we draw turns on the data-handling terms and the technical controls behind them (both necessary, neither alone sufficient). Second, we use 'data leakage' throughout in its neutral technical sense: the escape of data beyond the boundary within which it was meant to remain, whether or not anyone intended it. Third, privilege attaches to information, not to documents. A document is privileged to the extent that the information in it is privileged. In practice, however, this nuance often disappears because AI systems typically process whole documents. Unless material is anonymised or filtered before it reaches the system, privilege will be managed at document level, as will access rights. Where we frame our arguments as being about privileged documents, this is simply to better track the practical reality.

After setting out the doctrinal baseline (Section 2), the article proceeds from memorisation (Section 3), context window (Section 4) and RAG (Section 5) to the professional stakes associated with data risks posed by those three data storage and processing modes (Section 6). Section 7 concludes.

---

[3] *United States v Heppner* No 1:25-cr-00503-JSR (SDNY, 17 February 2026) (Rakoff J).

[4] Sir Colin Birss 'Legal Professional Privilege in the Age of AI' (Speech to the City of London Law Society, 22 April 2026), available at https://perma.cc/FX2Z-GL3Q.

[5] Eg, T Whittaker and S Bourton 'AI and Legal Privilege' (Burges Salmon, October 2025), available at https://perma.cc/M77Q-HMU9; P Fields 'When You Upload, What Have You Disclosed? AI, Confidentiality and the Embedded Document' (Barrister Magazine, 17 June 2026), available at https://perma.cc/VPR8-4TM5; J Perlin 'Client Confidentiality and Generative AI' (2027) 40 *Harv JL& Tech* (forthcoming), available at https://ssrn.com/abstract=6227838.

## 2 The doctrinal baseline

Our primary framework is that of legal professional privilege and its precondition: confidentiality. That said, the technical understanding of the GenAI data flows has a direct bearing on risk mitigation and information governance in respect of data protection, privacy, and professional responsibility (including ethics).

### 2.1 Privilege

Confidential communications between lawyers and clients attract special protection—legal professional privilege in England and Wales, attorney–client privilege in the United States, professional secrecy in civil law systems. The labels and doctrinal details vary, but the underlying principle is close to universal. In England and Wales, legal professional privilege is a fundamental right grounded, on one influential account, in the rule of law itself, and once established it is absolute: it cannot be overridden by some countervailing public interest in disclosure.[6] Privilege has two limbs: legal advice privilege protects confidential lawyer-client communications made for the purpose of giving or obtaining legal advice, understood broadly as the continuum of communications in the relevant legal context;[7] litigation privilege protects confidential communications with third parties whose dominant purpose is the conduct of litigation in reasonable contemplation.[8] Both extend to in-house lawyers as a matter of English law,[9] though not in EU competition investigations (a point to which we return).

Four features of the English doctrine frame the AI analysis. First, the courts have declined to extend privilege beyond the lawyer–client relationship by reference to the adviser's function rather than status: that is the lesson of *R (Prudential) v Special Commissioner of Income Tax*, where the Supreme Court refused to extend legal advice privilege to tax advice given by accountants.[10] Any argument that exchanges with an AI system giving 'legal advice' should themselves attract privilege must confront *Prudential* head on.

Second, confidentiality and privilege are related but structurally different protections, and the difference carries most of the analytical weight in the AI context. Confidentiality is a precondition of privilege: as Lord Scott put it in *Three Rivers (No 6)*, unless the communication for which privilege is sought is a confidential one, 'there can be no question of legal advice privilege arising'; the confidential character of the communication 'is not by itself enough to enable privilege to be claimed but is an essential requirement'.[11]

---

[6] *Three Rivers District Council v Governor and Company of the Bank of England (No 6)* [2005] 1 AC 610 (HL) at [34], per Lord Scott, crediting Professor Zuckerman's rule-of-law rationale. On the absolute character of the privilege once established, see *R v Derby Magistrates' Court, ex p B* [1996] AC 487 (HL).

[7] *Balabel v Air India* [1988] Ch 317 (CA), [1988] 2 WLR 1036.

[8] *Waugh v British Railways Board* [1980] AC 521 (HL); and see *R (on the application of Jet2.com Ltd) v Civil Aviation Authority* [2020] EWCA Civ 35, [2020] 2 QB 1027 (consolidating the dominant purpose jurisprudence); *Serious Fraud Office v Eurasian Natural Resources Corpn Ltd* [2018] EWCA Civ 2006, [2019] 1 WLR 791 (on reasonable contemplation of litigation).

[9] *Alfred Crompton Amusement Machines Ltd v Customs and Excise Commissioners (No 2)* [1972] 2 QB 102 (CA) at 129, per Lord Denning MR; affd [1974] AC 405 (HL).

[10] *R (Prudential plc) v Special Commissioner of Income Tax* [2013] UKSC 1, [2013] 2 AC 185.

All privileged communications are therefore confidential. But the converse does not hold, and the two protections operate differently. Privilege is a shield against *compulsion*: an exception to the general obligation to produce documents or answer questions, admitting no balancing against competing interests.[12] Confidentiality, by contrast, is a right to restrain *voluntary* use or disclosure. [13] It is broader in coverage but heavily qualified, since it yields to compulsory process and may yield to the public interest.[14]

Within that architecture, two doctrinally distinct routes can defeat the privilege: *loss of confidentiality*, which destroys the precondition (once information is genuinely in the public domain, there is nothing left for privilege to bite on), and *waiver*, ie, conduct of the privilege-holder treating the material in a manner inconsistent with maintaining the privilege. The two are sometimes conflated (including, as we shall see, in the first GenAI-related UK decision), despite having different consequences: loss of confidentiality is total, whereas waiver can be partial and purpose-limited.

Third, and crucially for AI systems, limited disclosure to a third party under obligations of confidence does not, without more, destroy either confidentiality or privilege. Privileged material is routinely placed with translators, e-disclosure providers and cloud hosts without anyone suggesting that privilege is thereby lost. The Privy Council has confirmed that even deliberate disclosure for a limited purpose, on terms preserving confidentiality, is not a general waiver.[15] The High Court then stated that communications of the substance of privileged advice to third parties, confidentiality being preserved, likewise remain protected.[16]

This line of authority strongly supports the proposition that processing privileged data through an enterprise-grade provider's infrastructure should not in itself imperil privilege. The question in every case is whether the terms and the technical reality genuinely preserve confidentiality, which is why understanding the three modes of data storage and processing is essential.

Fourth, even where privileged material escapes, the law does not always treat the protection as irretrievably 'gone'. It is the rights arising from the relationship of confidence, not privilege itself, that entitle a client to restrain third-party interference,[17] and in litigation the court's permission is required before inadvertently inspected privileged material may be deployed.[18] The remedy is imperfect—an injunction cannot claw back data that has been memorised by a model (on which more shortly)—but the absolutist language of some early commentary ('the privilege is gone') overstates the doctrinal position.

---

[11] *Three Rivers (No 6)*, *supra* n 6 at [24] per Lord Scott.

[12] *PCCW-HKT Telephone Ltd v Aitken* (2009) 12 HKCFAR 114 at [59], per Lord Hoffmann NPJ, quoted in C Passmore *Privilege* (5th edn, Sweet & Maxwell, 2024) para 1-415.

[13] *Price Waterhouse v BCCI Holdings (Luxembourg) SA* [1992] BCLC 583, per Millett J, quoted in Passmore, *supra* n 12, para 1-420.

[14] *Attorney General v Jamaican Bar Association* [2023] UKPC 6, per Lords Briggs and Hamblen, quoted in Passmore, *supra* n 12, para 1-416, citing *R v Derby Magistrates' Court, ex p B* [1996] AC 487.

[15] *B v Auckland District Law Society* [2003] UKPC 38, [2003] 2 AC 736 at [68]–[71].

[16] *USP Strategies plc v London General Holdings Ltd* [2004] EWHC 373 (Ch) at [9]–[21], per Mann J.

## 2.2 The confidentiality precondition

As will become clear, the status of privilege across different GenAI workflows turns largely on the confidential status of the processed data. The authorities offer different ways to test that precondition. The first asks whether there was, from the AI system *user's* standpoint, a reasonable expectation of privacy as against the party opposing the privilege claim.[19] Yet privacy and confidentiality are distinct constructs, so the better course is probably to test confidentiality directly on the footing established in *Coco*: whether the information was imparted in circumstances importing an obligation of confidence.[20] That enquiry is objective and conducted from the recipient's standpoint—whether a reasonable person in the recipient's shoes would have realised the information was given in confidence—and is intensely fact-specific. Applied to generative AI, the question is not what the user hoped, but what a reasonable provider would have understood from the circumstances of receipt. Provider terms are then the dominant circumstance.

Importantly, the precondition of privilege is that the material is confidential as against the person against whom the privilege is asserted. As Lewison LJ put it in *Brake v Guy*, whether it would have been confidential as against anyone else 'is irrelevant'.[21] The confidentiality enquiry in GenAI workflows will therefore often focus on whether the information is in truth in the public domain, accessible to the world at large.[22] Where material has become genuinely and extensively available, the courts accept that confidentiality has gone, however that happened, although in *Wee Shuo Woon v HT SRL* the Singapore Court of Appeal held that privileged emails stolen by hackers and published on WikiLeaks had *not* entered the public domain, because what matters is not only how accessible material is but how far it has in fact been accessed.[23]

---

[17] Passmore, *Privilege* (5th edn) paras 1-417 and 1-421, discussing *Lachaux v Independent Print Ltd* [2015] EWHC 3677 (QB), affd [2017] EWCA Civ 1327.

[18] CPR 31.20; *Al Fayed v Commissioner of Police of the Metropolis* [2002] EWCA Civ 780 at [16], setting out the governing principles.

[19] *Simpkin v Berkeley Group Holdings plc* [2017] EWHC 1472 (QB), [2017] 4 WLR 116 at [31]; *Tchenguiz v Imerman* [2010] EWCA Civ 908.

[20] *Jinxin Inc v Aser Media Pte Ltd* [2022] EWHC 2856 (Comm), [2023] 1 WLR 1084 at [37]–[38] and [40], per Simon Salzedo KC, applying *Coco v A N Clark (Engineers) Ltd* [1968] FSR 415 at 419 and 420–21, per Megarry J; and see Passmore, *Privilege* (5th edn) paras 2-092 and 2-103ff.

[21] *Simpkin v Berkeley Group Holdings plc*, *supra* n 19 at [31]–[35], per Garnham J, approved in *Brake v Guy* [2022] EWCA Civ 235 at [78], per Lewison LJ. Contrast *Jinxin* at [59], where provisions permitting the monitoring of corporate email were held 'commonplace in modern corporate life' and did not displace confidentiality.

[22] *Attorney General v Guardian Newspapers Ltd (No 2)* [1990] 1 AC 109 (HL) (*Spycatcher*) at 282 (Lord Goff).

Which test is applied dictates what a party must prove. On reasonable expectation, the enquiry runs to the prompts, the interface, the sign-up terms and the user's own understanding. On obligation of confidence, it is contractual and documentary: terms of service, data processing agreement, sub-processing chain. On public availability, it is technical and empirical: can a third party in fact retrieve the material, and by what means?

# 3 Memorisation: The model's permanent memory

## 3.1 How data becomes part of the model

Memorisation is the most permanent (and the most familiar) mode of data storage and processing in GenAI. When a large language model (LLM) is trained, the training algorithm processes hundreds of billions, and now trillions, of words—books, websites, code, academic papers, legal documents. Statistical patterns in that material are encoded ('baked') into the model's parameters that form hundreds of billions of numerical values constituting its long-term memory. Technically, parameters are not computer memory in the traditional sense (a storage space for writing and reading information), although information is, loosely, 'written' into model parameters during training and 'read' during inference. Crucially, what is baked in is not only general linguistic patterns. Specific, verbatim content can be encoded in non-trivial ways across those parameters, and extracted again with the right prompting techniques. As a result, the memorised sequence of words may be disclosed to other users of that model. Computer science researchers have repeatedly demonstrated that LLMs can 'memorise' parts of their training data in this way and can be induced to emit them verbatim, including personally identifiable information.[24]

The best-known real-world example comes from a different industry. In early 2023, Samsung Electronics lifted an internal ban on ChatGPT. Within weeks, three separate incidents had occurred in which engineers reportedly pasted confidential material—including proprietary source code and internal meeting notes—into the tool.[25] Under OpenAI's terms applicable to consumer products at the time, that data could be used to train OpenAI's models.[26] Samsung engineers' input was therefore eligible for training use, and so capable of being memorised and reproduced verbatim by a publicly accessible model.

---

[23] *Wee Shuo Woon v HT SRL* [2017] SGCA 23 at [36]–[37] and [43]; contrast *Glencore International AG v Commissioner of Taxation* [2019] HCA 26. See Passmore, *Privilege* (5th edn) paras 7-041 and 7-050ff.

[24] N Carlini and others 'Extracting Training Data from Large Language Models' in *Proceedings of the 30th USENIX Security Symposium* (2021); J Hayes and others, 'Measuring Memorization in Language Models via Probabilistic Extraction' in *Proceedings of the 2025 Conference of the Nations of the Americas Chapter of the Association for Computational Linguistics: Human Language Technologies (Vol 1)* (2025).

[25] 'Concerns become reality… Samsung Electronics "misuse" emerges one after another as soon as the ChatGPT ban is lifted' *The Economist* (Korea edition) (30 March 2023), available at https://perma.cc/3UL4-9CUV.

[26] Cf OpenAI 'Privacy Policy', openai.com/policies/privacy-policy. As of June 2026, consumer-tier inputs may be used for model training subject to opt-out; API and enterprise-tier inputs are excluded from training by default.

The lesson is obvious: no confidential data should be pasted into a consumer-grade tool whose terms of service generally permit input data to be used for training unless the user opts out. Importantly, even under active opt-out settings, model providers can treat in-session feedback (when a user rates the AI response as good or bad) as an 'opt-in' choice for the whole session. Some providers may also store and reuse user prompts if those are flagged classified by safety-oriented algorithms as being in violation of the terms (eg, if you ask for a bomb recipe).

## 3.2 *Munir*: Consumer-grade tools and waiver

Memorisation was also used in *Munir*[27] as the broad technical basis for the argument that feeding data into consumer-grade tools is like putting it in the public domain, but that view is imprecise. In November 2025, the Upper Tribunal (Immigration and Asylum Chamber) addressed in *Munir* for the first time in England and Wales what happens to privilege when privileged material is uploaded to a consumer-grade AI tool. *Munir* began as two conjoined proceedings concerning hallucinated citations. No privilege dispute was before the Tribunal, no party argued the point, and the relevant passage cites no authority. The privilege statement in *Munir* is therefore obiter, although the Tribunal elevated the warning into its formal headnote:

> Uploading confidential documents into an open-source AI tool, such as ChatGPT, is to place this information on the internet in the public domain, and thus to breach client confidentiality and waive legal privilege, and any such conduct might itself warrant referral to the regulatory body and should, in any event, be referred to the Information Commissioner's Office.[28]

On this view, privilege is not weakened or qualified by the upload. Instead it is gone. The Tribunal contrasted what it called 'open-source' tools, such as the free version of ChatGPT, with 'closed source AI tools which do not place information in the public domain, such as Microsoft Copilot', said to be available for tasks such as summarising 'without these risks'.[29]

The Tribunal's analysis is, with respect, doubly imprecise. The terminology is technically misinformed: ChatGPT is not 'open-source' in any accepted sense—its weights and code are proprietary,[30] and so the memorised content is not published anywhere (not even in the form of published model parameters). It is thus misleading to say, as the updated judicial guidance does, that '[a]ny information that you input into a public AI chatbot should be seen as being published to all the world'.[31] Whether the memorised data can be characterised as publicly available depends on who else can use the same model and how easily they could retrieve the data by prompting. The European Data Protection Board (EDPB) has recognised the data protection significance of this point. In Opinion 28/2024 it concluded that AI models trained on personal data cannot automatically be regarded as anonymous; whether personal data can be extracted from a model must be assessed case by case, having regard (among other things) to the likelihood of regurgitation of training data.[32] In other words, data protection regulators treat the model itself as potentially containing personal data and this view maps onto the confidentiality analysis.

---

[27] *Supra* n 1.

[28] *Ibid*, headnote para 4, see also [21] and [60].

[29] *Ibid* at [21].

[30] Open Source Initiative 'The Open Source AI Definition 1.0' (28 October 2024), available at https://perma.cc/AXA6-EQRM, requiring that the data information, the complete training and inference code, and the model parameters all be released on OSI-approved terms. On that definition ChatGPT's models are not open source.

More importantly, the legal characterisation in *Munir* conflates the two doctrinal routes distinguished above. Material pasted into a consumer-grade tool is not literally placed 'on the internet in the public domain': it is disclosed to an identified provider on terms permitting retention, training use and onward disclosure. On the *Spycatcher* test it is not thereby so generally accessible that it cannot be regarded as confidential, and the courts have declined to find a loss of confidentiality even where privileged emails had in fact been uploaded to the internet by hackers.[33] The better analysis is not loss of confidentiality through publication but that disclosure on such terms is inconsistent with maintaining confidence as against the provider, and so defeats the confidentiality (even if analysed as *reasonable expectation*) on which privilege depends (or, where the client authorises the upload, amounts to waiver). On the limited-disclosure authorities an upload to an enterprise-grade deployment under confidentiality obligations sits on the same side of the line as the e-disclosure provider or the cloud host, and should not in itself defeat privilege.[34]

Commentators have likewise observed that the Tribunal's blunt public-versus-private line is too absolute.[35] We agree: the question is not the label a system carries but whether the user can reasonably expect confidentiality given the contractual and technical reality of how the data is in fact handled. If a platform genuinely prevents external access to inputs, retention, and retraining on those inputs, using it should not in itself defeat privilege.

---

[31] Courts and Tribunals Judiciary 'Artificial Intelligence (AI) Guidance for Judicial Office Holders' (31 October 2025) at 4, available at https://perma.cc/5NEM-U57H.

[32] European Data Protection Board 'Opinion 28/2024 on certain data protection aspects related to the processing of personal data in the context of AI models' (adopted 17 December 2024), esp paras 29–58.

[33] *Supra* n 22 and n 23.

[34] *B v Auckland*, *supra* n 15.

[35] See (eg) G Johnston 'Legal Privilege and AI under English Law' (Juralio, 8 March 2026), available at https://perma.cc/V4FF-ARUN; T Whittaker and B Kolodziej 'AI and privilege' (Burges Salmon, 13 April 2026), available at https://perma.cc/8A55-U4V4.

### 3.3 *Heppner*: The terms of use decide

In February 2026, the Southern District of New York reached a position similar to that in *Munir*, though by a different route. In *United States v Heppner*,[36] the defendant, already under federal investigation, used the consumer-grade tier of Anthropic's Claude chatbot (on his own initiative, not at his lawyers' direction) to generate documents outlining his defence strategy, which he shared with his lawyers. After his arrest, the FBI seized some 31 such documents, over which he asserted privilege.

Judge Rakoff held that the documents were protected neither by attorney–client privilege nor by the work product doctrine. Three strands of the reasoning matter here. First, the communications were not between client and attorney: 'Because Claude is not an attorney, … that alone disposes of Heppner's claim of privilege' (at p 5). Second—and most instructive here—there was no reasonable expectation of confidentiality, because the consumer terms the defendant had accepted permitted the provider to collect inputs and outputs, use them for training, and disclose them to third parties including governmental authorities. Third, work-product protection failed because counsel had neither prepared nor directed the preparation of the documents and because material unprotected when created does not become protected by later reaching a lawyer.[37]

The analytical lesson echoes the critique of *Munir*: the 'open' versus 'closed' label is beside the point; what mattered were the terms the user had accepted, which defeated any reasonable expectation of confidentiality. American commentary has questioned the decision's categorical edge, noting that courts do not treat the use of Gmail or cloud storage as destroying confidentiality even though those providers can access user content.[38] But we believe this commentary misses the key reason why memorisation can defeat confidentiality, ie, that it allows the data to be accessed by other users of the model too.

### 3.4 Enterprise tools, fine-tuning and irreversibility

Setting these imprecisions aside, the underlying sentiment on both sides of the Atlantic is sound: enterprise-grade systems operating within a secure contractual framework (where the provider commits not to use customer data to train models available to other users[39]), and configured to match it, pose far fewer risks. Even so, enterprise-grade tools are not risk-free. Many legal organisations will want to fine-tune[40] a private instance of an AI model with the data in order to, for example, control running cost, boost performance, or attune the model to their preferences. Here the memorisation risk is more subtle. It is less about whether the data enters a publicly accessible model—in this scenario it should not—and more about irreversibility.

---

[36] *Supra* n 3.

[37] *Ibid* at pp 4–10 of the slip opinion.

[38] E X Guo '*United States v Heppner*' (Harvard Law Review Blog, 23 March 2026), available at https://perma.cc/CML7-S4N9.

[39] See (eg) OpenAI, openai.com/enterprise-privacy/ (ChatGPT Enterprise commitments); Anthropic 'Privacy Policy', anthropic.com/legal/privacy and data processing practices disclosed via Anthropic Trust Center, https://trust.anthropic.com/.

[40] Fine-tuning is the process of further training a pre-trained model on a smaller, task- or domain-specific dataset to adapt model parameters toward a particular objective while retaining the general capabilities acquired during pre-training.

Once a model is pre-trained or fine-tuned on specific data (say private or client-specific data), that information cannot be selectively removed because the information is distributed across the model parameters. There is no 'delete this data' button on a trained AI model and an enterprise-grade contract with the provider does not change that. This could harm the data subjects and violate their right to erasure under the UK and EU General Data Protection Regulation (GDPR).[41] 'Machine unlearning' remains an unsolved research problem at production scale.[42] And one must not lose sight of the prior question: a valid legal basis is needed for using client data, including personal data, to train an AI model in the first place.

Two consequences deserve emphasis, because they factually affect the confidentiality question and resist traditional data governance frameworks. First, a privately fine-tuned model is itself a new repository of confidential information: fine-tuning on the matter files of several clients irreversibly commingles their confidential information, and where conventional information barriers assume that documents (and the confidential information contained therein) can be segregated, a fine-tuned model cannot technically instantiate such segregation. Second, the assessment is temporal: a model fine-tuned today will be judged against the extraction techniques of tomorrow, if it remains in service, or in backups, for years.

This bears on a question many firms and in-house teams are considering: should they fine-tune a private model on their own precedents, matter files and house style? In most cases, not yet. Most of the benefit sought from fine-tuning can be substantially achieved through in-context learning instead, by giving the model what it needs at the point of use rather than baking it into the model and risking the collapse of information barriers. How that works is the subject of the next section.

The memorisation risk therefore reduces to two practical rules: know the tool's training policy before using it on client or sensitive work; and hold off fine-tuning on client or personal data until erasure has a workable technical solution. These maxims are consistent with the position now taken by the professional bodies. The Law Society's guidance says that no confidential data should be put into free online generative AI services, and both the Bar Standards Board (BSB) and SRA identify training-data transfer as a particular threat to confidentiality and privilege.[43]

---

[41] UK GDPR, art 17; Regulation (EU) 2016/679, art 17.

[42] Eg, S Liu and others 'Rethinking machine unlearning for large language models' (2025) 7 *Nat Mach Intell* 181; G-O Barbulescu and P Triantafillou 'To Each (Textual Sequence) Its Own: Improving Memorized-Data Unlearning in Large Language Models' in *Proceedings of the 41st International Conference on Machine Learning* (2024) PMLR 235:3003-3023.

[43] The Law Society 'Generative AI – the essentials' (guidance, updated September 2025), available at https://perma.cc/9UMX-9RGQ; SRA 'Risk Outlook report: the use of artificial intelligence in the legal market' (20 November 2023), available at https://perma.cc/C5XS-ZCN5; Bar Standards Board 'Guidance on the use of Artificial Intelligence and Other Technologies' (current from 18 May 2026), available at https://perma.cc/89P6-GNCQ, section 4 and rC86.3 commentary.

Lawyers who bar consumer-grade tools from client work, and client data from AI training, are applying the same logic.

# 4 The context window: In-session working memory

## 4.1 In-context learning is not training

If memorisation is the model's permanent memory, baked in at training time, the context window is its short-term working memory, operating at inference time—during the live session. The ability of an LLM-based system to use data supplied during the live session is often called 'in-context learning'. This may easily confuse lawyers reading vendor documentation. 'In-context learning' is not machine *learning* in the training sense, because the model's parameters do not change. 'In-context learning' simply means giving the model what it needs for the task within the conversation itself—a contract for review, background facts for a summary, or a policy document for analysis. The model works with what it is given for that session and produces an output from this temporarily 'learned' context. Nothing is permanently baked in, and when the session ends the model itself retains nothing.

In-context 'learning' is useful for lawyers who need the model to consider all kinds of relevant documents when producing its answer, and in many ways it is much safer than fine-tuning, because the data does not persist in the model. But even if the model retains nothing, at least three categories of risk arise: disclosure to the model provider; shared and persistent context; and prompt injection. Before turning to those, one practical limitation deserves attention: the size of context windows and the long-input problem.

## 4.2 Context rot and the long-input problem

Context windows are measured in tokens (a token is roughly three-quarters of a word). Modern windows can be large—typically several hundred thousand tokens (as of 2026), ie, the length of a book, or the documentary material for a simple piece of litigation—but they are not infinite.[44] What happens when the uploaded material is bigger than the window? Some tools refuse the upload and report the file is too large. That is the transparent response. Others process the input in batches or compress it, imposing an artificial dividing line, or even content reformulation, on the uploaded material. Yet others silently truncate the input: they process the first part and quietly ignore the rest. The model then answers confidently from an incomplete or skewed picture, with no flag that it never processed the whole file, or that it had modified the uploaded material. Figure 2 illustrates the failure mode.

[44] The figures are indicative for typical professional deployments in 2026; advertised maxima vary by model and tier, and effective usable context is generally smaller than the advertised limit.

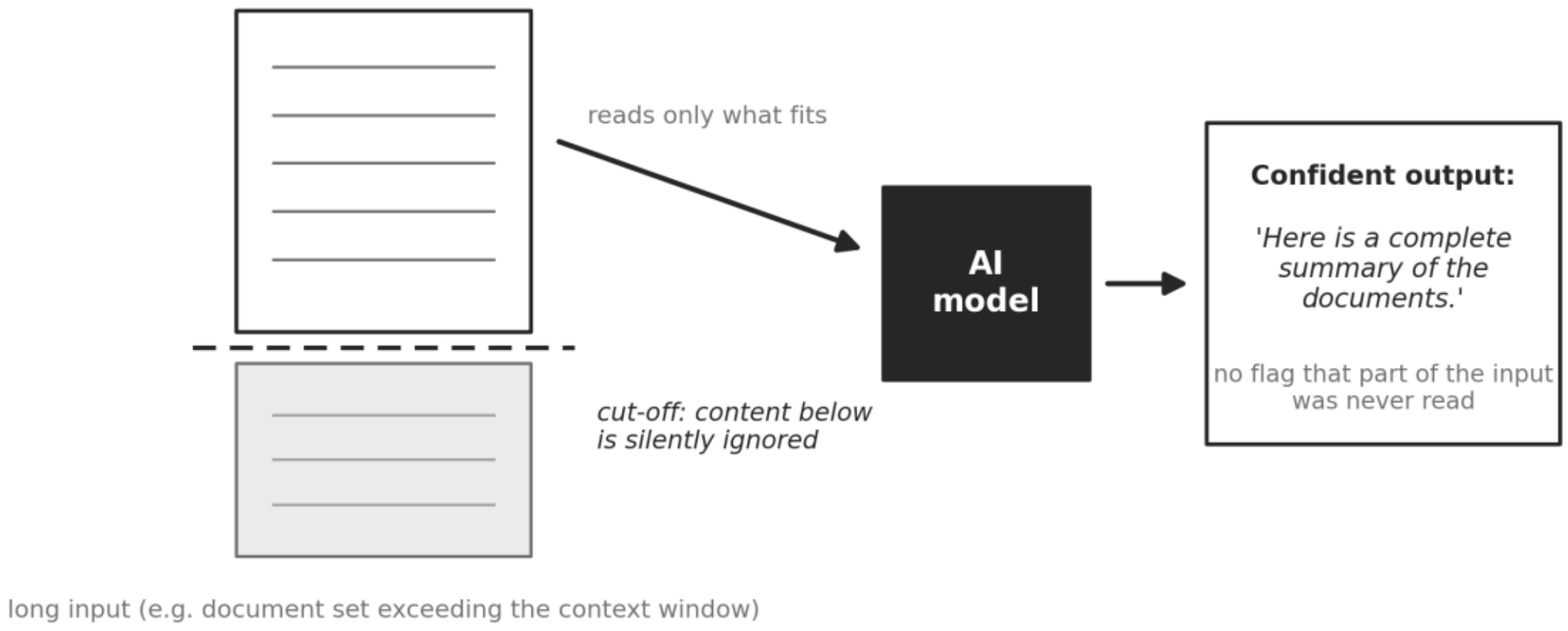

*Figure 2: Silent truncation of long inputs*

The consequences are serious for legal tasks built on long documents. Imagine a contract review that misses the final quarter, or a risk summary that fails to flag a clause buried near the end. Industry experts call this family of failures 'context rot': the quiet degradation of a model's performance and reliability as inputs grow longer or the window fills up.[45] The phenomenon is broader than truncation. Models attend less reliably to material in the middle of long inputs than at either end (positional bias in long contexts);[46] reasoning quality degrades as input volume grows even where the relevant information is present; and compression to fit the window may strip the very language on which legal meaning depends.[47] The output looks as polished as ever, but without verification one cannot trust that it rests on the whole document.

The danger for practice is that this failure mode is invisible to traditional supervisory review. The error resembles nothing a supervising partner has been trained to catch, yet it can compromise the output. And silent pre-processing to fit the window may itself pose confidentiality risks, depending on what technical methods are used to achieve the compression goal.

The practical responses are straightforward. Absent satisfactory answers from the provider, treat any submission beyond roughly half the advertised window length as long, and test for degradation by asking a question whose answer lies in the closing pages. For very large document sets, RAG (discussed in Section 5) may be a safer architecture than squeezing everything into one context window. Technologists are working on a general solution to the context window limitations, but the limitations are far from resolved.[48]

[45] H Wu 'Silent but Deadly – Context Rot Problems in Legal' (Syntheia, 27 March 2026), available at https://perma.cc/TY64-EUBF; K Hong and others 'Context Rot: How Increasing Input Tokens Impacts LLM Performance' (Chroma Technical Report, 14 July 2025), available at https://perma.cc/43YF-DP88.

[46] N F Liu and others 'Lost in the Middle: How Language Models Use Long Contexts' (2024) 12 *Transactions of the Association for Computational Linguistics* 157.

[47] *Ibid* (positional degradation); M Elaraby and D Litman 'ARC: Argument Representation and Coverage Analysis for Zero-Shot Long Document Summarization with Instruction Following LLMs' in *Proceedings of the 19th Conference of the European Chapter of the Association for Computational Linguistics (Vol 1)* (2026) (demonstration of the issues on legal data: CanLII); H Jiang and others 'LLMLingua: Compressing Prompts for Accelerated Inference of Large Language Models' in *Proceedings of the 2023 Conference on Empirical Methods in Natural Language Processing* (2023).

## 4.3 Disclosure to the model provider

Let us turn to the confidentiality risks proper. Everything put into the context window is visible to the model and shapes its outputs during the session. When the session ends, the window is cleared. That is the theory. In practice, the data can travel further than users may expect.

Every major consumer-grade AI tool routes user input through the provider's infrastructure. So do the leading legal AI platforms: at the time of writing, prominent legal tools such as Harvey and Legora sub-process data through the major model providers—Harvey through OpenAI and others, Legora through Azure-hosted OpenAI models and others.[49] Leading vendors have built their offerings around the contractual architecture we describe: Harvey guarantees that customer inputs, outputs and uploaded documents are not used to train the underlying models, requires zero data retention from its upstream model providers, and offers regional data residency, audit logs and customer-set retention policies; Legora makes equivalent no-training and no-retention commitments, holds an ISO 42001 AI-governance certification (as does Harvey), and provides configurable retention with deletion of all data at contract end.[50] These commitments are evidential rather than dispositive, and not interchangeable: each goes to whether the material reached the provider in circumstances importing an obligation of confidence. None is a silver bullet, and certification of a provider's AI management system says nothing about whether a given deployment excludes training use, for example.

Now the key point is that unless the firm or the client runs its own sealed-off, locally hosted instance, the data must go elsewhere to be processed. The prompt, and with it the whole context window, passes through a model that usually sits on a third party's infrastructure.

This technical reality does not change the legal test and so this form of disclosure alone is not a problem for privilege, provided there is confidentiality all along the way. To see whether and when disclosure of data to the model provider (via context window) destroys confidentiality (or precludes privilege from arising), it helps to separate four scenarios that are often run together.

---

[48] A Modarressi and others 'NoLiMa: Long-Context Evaluation Beyond Literal Matching' in *Proceedings of the 42nd International Conference on Machine Learning* (2025) PMLR 267:44554-44570; J Su and others 'RoFormer: Enhanced Transformer with Rotary Position Embedding' (2024) 568 *Neurocomputing* 127063; D Aumiller and others 'Structural Text Segmentation of Legal Documents' in *Proceedings of the Eighteenth International Conference on Artificial Intelligence and Law* (2021) 2.

[49] See (eg) Microsoft 'How Legora is transforming the legal workspace using Azure OpenAI Service' (Microsoft customer stories, 27 May 2025), available at https://perma.cc/5X6Y-2SFC; Legora 'Trust Center' (security.legora.com/); Harvey 'Security' and 'Trust Center' (harvey.ai/security and trust.harvey.ai/).

[50] Harvey 'Your Data, Your Control: How Harvey Manages Customer Data' (20 March 2026), available at https://perma.cc/J9FE-BQR9; Legora 'Security and compliance' (legora.com/security). Both platforms publish SOC 2, ISO 27001 and ISO 42001 attestations, for example.

In the first, a litigant consults a consumer-grade AI tool directly for what is functionally legal advice, with no lawyer involved. The Chancellor of the High Court has described typing a consumer-law question into the public version of ChatGPT and receiving a structured analysis ending with a request for further instructions.[51] Such an exchange may look like one between client and adviser, but no privilege attaches, because the 'adviser' is not a legal professional, and *Prudential* bars any functional extension.

In the second, a client uses an AI tool on their own initiative *alongside* an existing retainer and feeds the results to their lawyer. That is *Heppner*, where material unprotected when created did not acquire protection by being routed to the lawyer later.

In the third, the lawyer uses an AI tool in the course of producing advice. Here the analysis is the most forgiving, provided the system is secure. As Sir Colin Birss put it, lawyers have always been entitled to consult textbooks and online resources without undermining privilege.[52]

In the fourth, a client uses an AI tool to review (or get a second opinion on) privileged legal advice. Here too the terms and technical guarantees are essential, since under the limited waiver principle material can be shared confidentially without waiving privilege more broadly.

In all four scenarios, the privilege analysis is not simply a re-run of the cloud computing debates of the 2010s. Privileged material has always passed through third-party infrastructure—postal services, email servers, hosted document management systems—and the profession has rightly proceeded on the basis that confidential transit through a provider bound by confidence does not destroy privilege.

Three features distinguish GenAI transit. First, under consumer-grade terms the provider may acquire *use* rights over the content—above all to train models on it—which no postal analogy parallels, and which converts transient transit into potentially permanent memorisation (Section 3).

Second, the systems do not merely store or carry the material but *process* it, so that the content shapes outputs, logs, safety-review queues and caches in ways that multiply the points of persistence. For example, even if data is not memorised, it will often be cached. Caching is a general computing technique: when a task is expensive or slow, the system quietly saves the result somewhere quick to access, so that it need not redo the work next time. This happens inside AI systems too, without any obvious notification. Caching also affects data in the context window—the input prompts and responses may be cached during an open session with the AI system in order to minimise computation cost. In consumer-grade tools, the provider has no commercial incentive to re-run full computations in order to produce AI-generated answers to the same (or substantially the same) prompts. In a poorly configured system, different users could therefore receive the same pre-computed answer, which may rest on another user's privileged or otherwise sensitive prompt.

---

[51] *Supra* n 4, paras 15–19.

[52] *Ibid*, para 26.

Third, the resulting artefacts—chat histories, cached responses, embeddings—are new data, generated partly or entirely outside the lawyer's systems and the client's control. Prompts and outputs are 'documents' for disclosure purposes, a document being anything in which information of any description is recorded.[53] Those produced by a lawyer using a secure system in the course of giving advice may well be privileged working papers, consistently with the Chancellor's analysis above; prompts written by a client or a non-lawyer in the business will usually enjoy no such protection. Document retention policies, litigation holds and disclosure protocols should therefore address context window data explicitly.

Whether these differences are decisive will thus depend on the terms in each case and on the relevant commitments holding, and being verifiable, down the chain. And this deserves a fresh analysis rather than a straightforward extension of the existing debates. The relevant passage in *Munir* is unreasoned obiter guidance from the Tribunal and there is thus ample scope for the position to develop.

What, then, does an enterprise-grade AI deployment with strict privacy controls require in practice? At minimum a data processing agreement identifying the provider as processor; contractual exclusion of inputs and outputs from model training; zero or strictly time-limited retention of session data, including server-side logs; no human review of inputs save in defined abuse scenarios; encryption in transit and at rest; and clarity about where, geographically and organisationally, the data is processed—including whether the legal AI vendor sub-processes through an upstream model provider, and on what terms. These questions should be asked, and the answers documented, before any client material enters the system. If a challenge comes, the firm must be able to evidence that its expectation of confidentiality was reasonable at the time of use.

That all being said, the data exposure through the context window is still risky in case of compulsion directed at the provider, rather than the law firm. The privilege/confidence architecture set out above explains why a provider's contractual duty of confidence protects only against voluntary disclosure. It is no answer to compulsory process directed at the provider—a regulator's statutory information demand, a foreign court's production order, or a litigation preservation order over user session data (eg, all context window history) in proceedings to which the law firm is a stranger.[54] There, the only protection capable of resisting compulsion is the client's privilege itself. That is perhaps the strongest reason for ensuring privilege survives transit through the tool: if confidentiality has been preserved, the client has something to assert against the compelled provider; if not, contractual confidence alone is unlikely to hold the line.[55]

---

[53] CPR 31.4; and see Practice Direction 57AD, para 2.5 for proceedings in the Business and Property Courts.

[54] *Price Waterhouse v BCCI*, *supra* n 13. On the limited statutory carve-outs protecting privilege against regulators' information-gathering powers, see Passmore, *supra* n 12, para 1-413.

[55] Cf *R (McKenzie) v Director of the Serious Fraud Office* [2016] EWHC 102 (Admin), [2016] 1 WLR 1308, discussed in Passmore, *supra* n 12, para 1-418: where material is taken under coercive powers, the safeguards required of the holder differ from those of a fiduciary—underscoring that the client's strongest protection in third-party hands is the subsistence of the privilege itself.

Commentary likewise argues that a firm cannot discharge its confidentiality obligations on vendor assurances alone. Even where training is contractually excluded, sub-processor chains and internal review teams remain disclosure routes, inputs may be logged for troubleshooting, cross-border processing creates exposure to foreign compulsory powers, and derivative information may persist in forms difficult to audit.[56] We agree: the vendor commitments described above are real progress, but they are the beginning of due diligence, not its end.

### 4.4 Shared and persistent context

Another context window risk arises from the ability of some GenAI tools to allow context to persist across sessions and to be shared with colleagues—project-based workspaces in the leading assistants are current examples.[57] This is useful since the model can retain matter background, hold a running brief and apply consistent drafting guidelines across a team. But confidential client data entering the workspace may be accessible to everyone in it.

The unintuitive part is that *what* is shared is not only uploaded documents but the shared context itself, including snippets of users' chats within the workspace. A lawyer who works through a sensitive aspect of a matter with an AI assistant inside a shared project may be disclosing that exchange to every member, including those who join later.

The conflict-of-interest and confidentiality implications are immediate, and English law sets the bar high. In *Bolkiah v KPMG* the House of Lords held that a firm holding a former client's confidential information and proposing to act in a matter adverse to that client must show that there is no real risk of the information reaching those acting on the other matter; ad hoc arrangements improvised within a single department will not do—an effective information barrier must be 'an established part of the organisational structure of the firm'.[58]

A shared AI workspace is the antithesis of an established structural barrier: it is a deliberately constructed channel through which matter-related context flows to everyone with membership. Barriers engineered at the level of file permissions can be bypassed entirely by a shared context window sitting outside them, and on *Bolkiah* principles it would be for the firm to show that no real risk of leakage through that channel exists. Anyone contributing to such a workspace should be able to answer who else can see the context they are working in. Workspace membership should be provisioned from the same source of truth as matter-level access controls.

### 4.5 Prompt injection

The final context window risk is prompt injection: content embedded in material the AI is invoked to analyse, containing hidden instructions designed to hijack the model's processing. The example most relevant to practice is perhaps a counterparty sending a contract for review. Embedded invisibly in the file—white text, a tiny font, document metadata, or a form not intelligible to humans at all—is an instruction directing the AI to summarise the agreement favourably, or to ignore a clause. A court has already withdrawn a litigant's electronic filing rights for concealing such instructions in his pleadings.[59] The danger is that the model processes all the input indiscriminately and generates output accordingly.

---

[56] E Dimitrova 'AI and Client Confidentiality: The next regulatory faultline' (Legal Futures, 27 February 2026), available at https://perma.cc/P5UK-DECJ.

[57] Eg, the AI 'project' environments or shared AI 'workspaces' which retain project files, instructions and chat-derived context across sessions and share them with project members.

[58] *Bolkiah v KPMG* [1999] 2 AC 222 (HL) at 239, per Lord Millett.

The UK National Cyber Security Centre (NCSC) has warned that prompt injection should not be analogised to familiar vulnerabilities such as SQL injection, and may be worse, because the underlying weakness cannot be fully patched. An LLM draws no inherent distinction between 'instructions' and 'data'—under the hood there is only ever the prediction of the next token—so injection cannot be mitigated in the way SQL injection can be through parameterised queries.[60] The NCSC's advice is to treat an LLM as an 'inherently confusable deputy' and to design systems so that the impact of a successful injection is contained, rather than hoping to prevent injection altogether. Where a tool is connected to other systems—email, document management, drafting tools—the question for the vendor is not whether it blocks prompt injection (none can, completely) but what the model or the 'AI agent' can *do* if an injection succeeds, and what deterministic safeguards limit it.

For practitioners the implications are twofold. Where documents come from external parties, AI-assisted analysis should be treated with scepticism and outputs verified against the source. Better still, the material should be cleaned or converted to a plain format to see if it contains any unexpected kinds of data before it enters the system. Many tools already apply automated screening of this kind, though none is perfect. Importantly, however, if a rogue actor gains access to confidential data regardless of any such reasonable defence measures, this data breach alone will probably not defeat confidentiality.[61]

## 5 Retrieval-augmented generation

### 5.1 The architecture

RAG is increasingly the architecture behind legal AI tools, and for good reason. Pre-trained LLMs have a knowledge cut-off date and have never been exposed to the firm's internal precedent bank or the client's files, so the model must be given relevant data at inference time. But, as Section 4.2 explained, the context window cannot hold everything at once.

RAG solves this with a separate, searchable knowledge store. Documents—cases, legislation, internal precedents, client files—are split into chunks of a few hundred words each, converted into numerical representations called embeddings, and stored in a vector database. When the user provides an input, the system retrieves into the context window the chunks most semantically similar (or otherwise relevant) to the query, and the model answers from both its trained knowledge and those chunks. The architecture reduces hallucination, sidesteps the long-input problem, and avoids baking client data into model parameters. But it creates a second place where client data persists, typically outside the document management system, and carries five specific risks. Figure 3, which is our own, shows the canonical pipeline and marks the points at which the risks discussed below arise (production systems vary, but the persistence points are common to the architectures in general use).

---

[59] *Elliott v New York Bariatric Group LLC* (Connecticut Superior Court, 6 August 2026, docket AAN-CV-25-6066141-S). See also V H Lantyer 'Prompt Injection in Court Filings: Generative AI in the Brazilian Judiciary, Algorithmic Procedural Bad Faith, and the Limits of Legal Sanction' (2026), available at https://ssrn.com/abstract=6762100.

[60] D Chismon 'Prompt injection is not SQL injection (it may be worse)' (National Cyber Security Centre blog, 8 December 2025), available at https://perma.cc/9EG2-YC3D.

[61] See text to n 23.

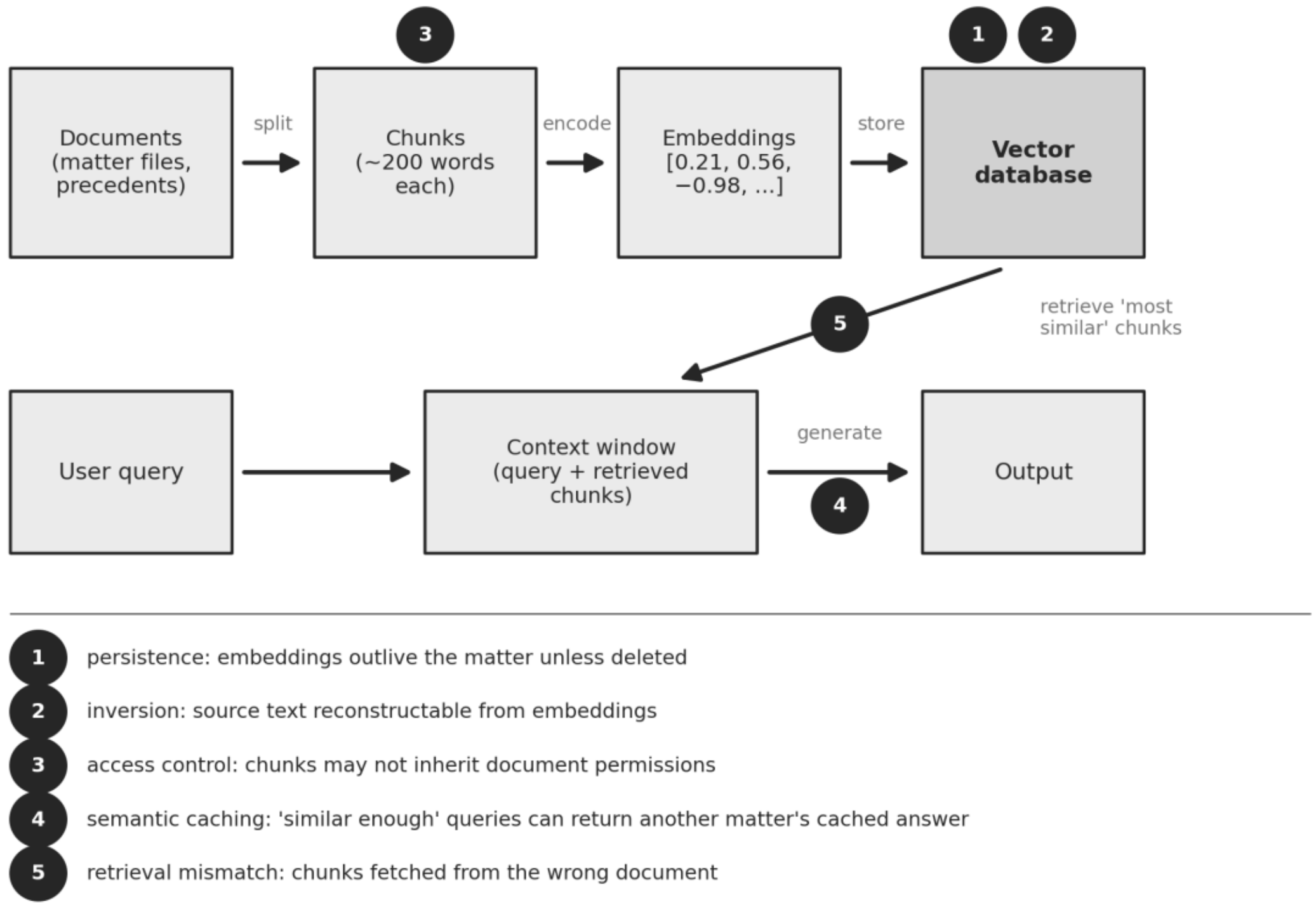

*Figure 3: A RAG pipeline and its principal risk points*

## 5.2 Embedding persistence

When a document is embedded and stored in a vector database, the data persists until explicitly deleted. This is the first risk. When a matter closes, a client relationship ends, or a document falls outside the retention policy, the embedded version does not by default go with it. The vector database is, in substance, a copy of the client file in another format, sitting outside the systems to which retention schedules are usually applied. Well-governed teams therefore set explicit retention policies for RAG knowledge bases (30 days is typical for matter-specific stores) and check whether retention is customer-configurable. In the leading tools the control typically exists, but it operates only if someone configures it.

That said, even a well-drafted retention policy brings complications. The team loses access to the knowledge base on expiry, so project timelines must account for it. The point matters all the more because GenAI tools increasingly create RAG knowledge bases automatically (eg, from large documents uploaded into a project workspace, or from chat history) since RAG is also a technique for controlling the number of tokens in the context window. Data can thus enter a persistent retrieval store without anyone having decided to put it there.

## 5.3 Embedding inversion: The database is the information

It is tempting to assume that embeddings (long lists of numbers) are a safely abstracted, effectively anonymised representation of the underlying text. Recent research shows otherwise. Morris and colleagues demonstrated that iterative 'inversion' methods can recover 92 per cent of short (32-token) passages exactly from their embeddings, and can extract verbatim personal information, such as patient names, from embeddings of clinical notes.[62]

Text embeddings can therefore reveal almost as much as the text itself, if one assumes access to the embedding vectors *and* to (or knowledge of) the embedding model used to create them. Since most production systems use a handful of widely available embedding models, and inversion methods are getting stronger, the risk of an attacker who reaches a vector database and reconstructs meaningful content from it must be managed too.

From a confidentiality standpoint, the conclusion is simple: the vector database should be treated as the underlying documents would be (as part of the organisation's system of record, subject to the same access controls, encryption standards and retention schedules as the document management system).

## 5.4 Semantic caching

The third risk is caching (first explained in Section 4.3), which in our experience few practitioners have heard of. RAG systems cache chunk embeddings and sometimes whole query responses. Of the several forms caching takes, the riskiest for confidentiality in RAG-enhanced workflows is semantic caching.

In a RAG system with semantic caching, two 'similar enough' questions can return the same cached answer rather than running the full retrieval process twice. But 'similar enough' in the eyes of a caching algorithm can cut across client and matter boundaries: one fee-earner's query about a force majeure clause could, in a poorly configured system, return a cached response generated from a different matter's privileged documents. That is data contamination with direct implications for confidentiality and professional conduct. On any pipeline used for client work, semantic caching should thus be disabled by default or strictly partitioned per matter or per user (though this may raise the implementation cost), and the question 'Is response caching shared across users or matters?' belongs in every vendor due diligence questionnaire.

---

[62] J X Morris and others 'Text Embeddings Reveal (Almost) As Much As Text' in *Proceedings of the 2023 Conference on Empirical Methods in Natural Language Processing* (2023); D Seputis and others 'Rethinking the Privacy of Text Embeddings: A Reproducibility Study of "Text Embeddings Reveal (Almost) As Much As Text"' in *Proceedings of the Nineteenth ACM Conference on Recommender Systems (RecSys '25)* (2025).

## 5.5 Access control

The fourth risk is access control. On document management systems, documents carry permissions and information barriers are enforced at file level. The question is whether the RAG system inherits those permissions, or whether the retrieval database cuts across them.

In theory chunks should inherit the permissions of their parent files. In practice, general-purpose RAG stacks do not do this by default. There is no standard mechanism for reading access rights from an uploaded file into a third-party tool's retrieval logic. A document that only three people may open can become, once embedded, retrievable in answer to anyone's query.

The leading platforms are addressing the gap though. Harvey's iManage partnership lets the AI read documents in place under the document management system's access controls, and Legora offers matter-level permissioning on the same logic.[63] Microsoft 365 Copilot now retrieves through the Microsoft Graph and surfaces only content the querying user may already see, so that retrieval inherits existing SharePoint and OneDrive permissions.[64] That makes the quality of the underlying permissioning the controlling variable, since a site opened to 'everyone in the organisation' years ago becomes, on the day Copilot is enabled, a retrieval source for the whole firm.

The direction of travel is right, but the firm must obtain positive assurance—by testing the deployed configuration, not by reading a datasheet—that permissions actually propagate into retrieval. The safest approach where permission inheritance is lacking is to have one RAG pipeline per matter or per user, populated only with documents the relevant people may use and share. It is more work, but it maps onto professional obligations, limits the blast radius of any leak and keeps retrieval focused.

## 5.6 Retrieval mismatch

The fifth risk stems from RAG accuracy, and it affects practitioners however carefully the other four are managed. Researchers have recently identified and quantified a failure mode they call Document-Level Retrieval Mismatch (DRM): a RAG system searching a large legal database retrieves content from an entirely incorrect source document.[65] Not the wrong paragraph—the wrong document. The failure is especially common in legal settings, where documents are structurally similar. Contracts, statutes and judgments all follow conventions and this resemblance compromises retrieval, causing the model to generate a confident, well-written answer from possibly the wrong source. The user would not know without checking.

---

[63] 'Harvey Announces Tech Partnership With iManage' (Artificial Lawyer, 5 June 2025), available at https://perma.cc/7S4B-RBDF; Legora 'Security and compliance', *supra* n 50; on matter-level permissioning in client-facing workspaces, 'Legora Launches Client Collaboration Portal' (Artificial Lawyer, 7 November 2025), available at https://perma.cc/3AJ4-EFVU.

[64] Microsoft 'Data, Privacy, and Security for Microsoft 365 Copilot' (Microsoft Learn, updated 18 August 2026): Copilot 'only surfaces organizational data to which individual users have at least view permissions', and prompts, responses and Graph data 'aren't used to train foundation LLMs'.

[65] M Reuter and others 'Towards Reliable Retrieval in RAG Systems for Large Legal Datasets' in *Proceedings of the Natural Legal Language Processing Workshop 2025* (2025).

The same research suggests design-level mitigations (enriching chunks with document-level summaries markedly reduces DRM), but for practitioners the lesson converges with the access control point. The more similar the documents in a knowledge base, the higher the risk of retrieval from the wrong one, so per-matter RAG is safer for privilege and more reliable on accuracy grounds too. And on outputs that matter, retrieved sources should be opened and read, not merely observed to exist.

# 6 The professional stakes

The risks canvassed above engage the core duties of the profession. We address, in turn, the regulatory framework, privilege governance, the professional negligence analysis, the cost objection, and how far the argument travels beyond England and Wales.

## 6.1 The regulatory framework

No SRA rule mentions generative AI, but the existing framework already reaches every risk described above, anchored in the duty of confidentiality. Paragraph 6.3 of the SRA Code of Conduct for Solicitors requires the solicitor to 'keep the affairs of current and former clients confidential unless disclosure is required or permitted by law or the client consents'; the same obligation binds firms, which must in addition have effective governance structures, systems and controls ensuring compliance.[66]

The duty is not exhausted by refraining from deliberate disclosure: as *Bolkiah* makes clear, the lawyer must take effective steps against any real risk of confidential information escaping or being misused.[67] Every data processing operation discussed here—training, context transit, shared workspaces, vector stores, caches—can breach that duty without any document being 'sent' to anyone in the conventional sense. Pasting a client letter into a consumer-grade chatbot is, on the regulator's own analysis, such a breach. The SRA's Risk Outlook report identifies as 'particular threats' to confidentiality and privilege a staff member 'using an online AI, such as ChatGPT, to answer a question on a client's case', confidential data 'being revealed when it is transferred to an AI provider for training', and 'the output from an AI system replicating confidential details from one case in its response to another'.[68] The Law Society is blunter: where a firm uses 'a free, online generative AI service ..., do not put any confidential data into the tool'.[69]

---

[66] SRA Code of Conduct for Solicitors, RELs and RFLs, para 6.3; SRA Code of Conduct for Firms, paras 2.1 and 6.3.

[67] *Supra* n 58 at 235–39.

[68] SRA, *supra* n 43, section on confidentiality and privacy.

[69] The Law Society, *supra* n 43, para 3.3.

Confidentiality is not the only relevant head. Competence and supervision (paras 3.2–3.5 of the Code for Solicitors) require the practitioner to understand the tools through which the service is delivered and to supervise work effected through them—the lesson drawn explicitly in *Munir*, where the failure was supervisory before it was technological.[70] The duty not to mislead the court or others (para 1.4) is engaged by unverified AI output, as *Ayinde* and the Law Society's September 2025 update emphasise.[71] Client communication obligations bear on whether clients should be told that GenAI is used on their matters, with the Law Society advising that expectations be discussed.[72] And any serious breach (an upload of privileged material to a consumer-grade tool may well be one) engages the duty to report promptly to the SRA (paras 7.7–7.8).

In *Munir*, a legal professional whose firm had used a generative AI tool to summarise client correspondence and Home Office decision letters recognised the upload as a breach, self-reported to his regulators, and was not referred further by the Tribunal (a referral would otherwise have been made).[73] The lesson is that prompt self-reporting materially influenced the outcome for the individual. Information governance protocols should therefore cover incident response as well as prevention, including reporting to the regulator and, where personal data is involved, to the Information Commissioner's Office (with notification to affected data subjects where the risk to the data subject is high).[74]

The Bar Standards Board's guidance on AI, current from May 2026, addresses the same ground in greater technical detail, analysing the use of AI tools as *outsourcing* under rC86 of the BSB Handbook—the barrister remains responsible for the work and must ensure the provider's terms and technical capability can maintain the confidentiality obligations of Core Duty 6—and states that 'it is unlikely that free-of-charge, publicly available AI systems fulfil these conditions'.[75] It places general-purpose tools without adequate data protections, and agentic features, in the highest risk band; expects record-keeping to extend to AI audit logs and prompt deletion policies; and (corroborating our shared-context analysis) warns that a chambers-level AI assistant may allow other barristers to 'view or infer sensitive client information', 'particularly if other barristers in your chambers are representing an opposing party'.[76] It also expects counsel to consider whether a client's or instructing solicitor's own AI use has compromised privilege, and to advise accordingly.[77]

---

[70] *Supra* n 1, headnote para 2 and [56]–[58].

[71] *R (Ayinde) v London Borough of Haringey; Al-Haroun v Qatar National Bank QPSC* [2025] EWHC 1383 (Admin), [2025] 1 WLR 5147; The Law Society, *supra* n 43.

[72] The Law Society, *supra* n 43, paras 1.4 and 3.1.

[73] *Supra* n 1 at [14], [16] and [19].

[74] SRA Code of Conduct for Solicitors, RELs and RFLs, paras 7.7–7.8; UK GDPR, arts 33 and 34.

[75] BSB, *supra* n 43, section 4 and rC86.3 commentary.

[76] *Ibid*, sections 1 and 4.

The refreshed judicial *AI Guidance* tells office holders to 'treat all public AI tools as being capable of making public anything entered into them', warns that the publicly available chatbots 'remember every question that you ask them' and that the information is 'available to be used to respond to queries from other users', and adds that even with chat history disabled 'it should be assumed that data entered is being disclosed'.[78] That is *Munir*'s technical premise stated more carefully: the guidance speaks of public tools, and of what should be *assumed*, not of what is legally established.

Two further observations are relevant. First, accountability is non-delegable. As the SRA puts it, the firm remains 'responsible and accountable for the outputs' of the AI it uses;[79] vendor assurances allocate commercial risk, not professional responsibility. More recently, the SRA has even warned that uploading confidential client material to open source AI tools (SRA's terminology was adopted from *Munir*), 'will likely breach client confidentiality and as a result, legal professional privilege may be permanently waived'.[80] Second, abstention is no safe harbour either. The SRA observes that 'the risk to firms might not come from adopting AI, but from failing to do so',[81] while the UK Jurisdiction Taskforce contemplates that a professional may be liable for failing to use AI where a competent member of the profession would have done so.[82] Professional regulation thus expects a calibrated adoption of AI tools under effective governance, and this is what our analysis in this article is designed to facilitate.

## 6.2 Privilege governance

As *Munir* and *Heppner* illustrate, where the data-handling terms destroy any reasonable expectation of confidentiality, privilege fails or is waived, and the client bears the loss. Conversely, as the Chancellor of the High Court has indicated, a lawyer's use of a demonstrably secure system to assist in giving advice should not imperil the client's privilege,[83] and on the limited-disclosure authorities an enterprise-grade deployment under genuine confidentiality obligations should fall on the safe side of the line. The difference is in governance and, if a challenge comes, evidence of it. The firm that can produce its data processing agreement, its zero-retention configuration and its vendor due diligence record is in a far stronger position than the firm that can produce only an invoice.

---

[77] *Ibid*, sections 1 (Core Duty 7) and 6.

[78] *Supra* n 31.

[79] SRA, *supra* n 43, section on accountability.

[80] *Supra* n 2.

[81] SRA, *supra* n 43, section on opportunities from AI.

[82] UK Jurisdiction Taskforce 'Legal Statement on Liability for AI Harms' (July 2026) 43–44.

[83] *Supra* n 4, para 26.

But protecting privilege in the age of AI goes beyond governance and technology. A governance policy no one understands well enough to apply is not governance at all. Responsible governance includes the AI literacy of the individual practitioner (or the client), because the technology shapes what is possible, governance what should be done, and AI literacy what is responsibly done. Additionally, good governance must cover deliberate non-use of AI. Where exposure would be intolerable whatever the legal position, or the client does not wish their data processed by AI, the answer will be a restricted list with named clients or matters for which no GenAI processing is permitted at all.

## 6.3 The professional negligence analysis

The standard of care expected of a solicitor is that of the reasonably competent practitioner professing the relevant skill,[84] and it evolves with the tools and knowledge available to the profession.

Compliance with common practice will not necessarily suffice because where a practice widely followed in the profession carries a foreseeable and readily avoidable risk, a court may condemn the practice itself.[85] A firm whose AI deployment mirrors its peers' is not thereby safe.

The standard is also temporal, judged by the knowledge reasonably available at the time of the alleged breach rather than with hindsight, and that cuts both ways. Conduct excusable in 2023, when these failure modes were obscure, may be inexcusable in 2026, now that the regulator, the Law Society, the NCSC and the judiciary have all published on them. A lawyer who deploys generative AI without understanding the data flows described above thus risks falling below the standard. Retention, caching, memorisation, context sharing, context rot, silent truncation and retrieval mismatch are now documented, foreseeable failure modes, and entering client data under a risky configuration, or relying without verification on confident-looking output, is culpable, as the growing line of hallucinated-citation cases shows.[86] That said, one cannot be blamed for failing to anticipate an extraction technique unknown at deployment, though one can be blamed for failing to reassess as the threat landscape moves.

It should not be assumed that every AI-related mishap will give rise to a claim in damages. On current scope-of-duty analysis, the question is whether the loss claimed falls within the risk the duty existed to guard against.[87] Where a solicitor's retainer includes the safekeeping of the client's confidential and privileged information (as it almost always will), the loss of privilege through processing in an inadequately vetted system falls squarely within that risk. Causation will often be straightforward (but for the upload, the material would have remained confidential and privileged), and the loss—inspection of strategy documents by an opponent, or a prosecutor, as in *Heppner*—is concrete and was avoidable at modest cost. Quantification is harder, since the value of lost privilege manifests in litigation outcomes and settlement positions, but that does not affect liability or disciplinary consequences. The accuracy failure modes feed a more familiar pattern (negligent advice or drafting based on an incomplete reading of the documents) where the question is whether the lawyer's verification practice met the standard.

---

[84] *Bolam v Friern Hospital Management Committee* [1957] 1 WLR 582; for solicitors, *Midland Bank Trust Co Ltd v Hett, Stubbs & Kemp* [1979] Ch 384, [1978] 3 WLR 167.

[85] *Edward Wong Finance Co Ltd v Johnson Stokes & Master* [1984] AC 296 (PC).

[86] *R (Ayinde)*, *supra* n 71; *Munir*, *supra* n 1.

[87] *Manchester Building Society v Grant Thornton UK LLP* [2021] UKSC 20, [2022] AC 783; *Meadows v Khan* [2021] UKSC 21, [2022] AC 852.

There is also a client-facing dimension. In *Heppner*, it was the *client's* self-directed use of a consumer-grade chatbot that generated the unprotected material. A client cannot be supervised into compliance, but a client can be warned. Retainer letters and engagement communications should now routinely advise clients not to discuss the matter with, or upload matter documents to, consumer-grade AI tools (just as clients have long been warned about discussing litigation on social media). The practitioner press makes the same point. Where clients upload advice, correspondence and drafts (eg, to test the advice or summarise it for colleagues) and where the resulting output is thus no longer the lawyer's communication, its status is doubtful even where that of the original advice was not.[88] Given how natural it has become to 'ask the chatbot', a failure to give the warning may itself come to look like a gap in competent client care. For barristers, the BSB guidance now expressly contemplates advising clients where their own AI use may have compromised privilege.[89]

### 6.4 The cost objection

A fair objection is that these recommendations are not free. The burden falls hardest on small firms, legal aid practices and thinly resourced in-house teams, since cheap consumer-grade tools benefit those who can least afford bespoke deployments (a point the Chancellor of the High Court acknowledged in noting the pro-access role of AI-assisted litigants in person).[90]

Three responses can be given. First, the core recommendations of this article (know the data-handling terms, keep client data out of consumer-grade tools, partition by matter, verify outputs, comply with restriction lists) cost attention rather than capital, and the Law Society's SME-focused guidance appears to rest on the same premises.[91] Second, the real comparison is between governance costs and breach costs. A single privilege waiver, regulatory referral or notifiable data breach will consume more partner time than taking the necessary governance steps. Third, the standard of care accommodates proportionate, documented risk decisions. What it does not accommodate is ignorance of where the data goes.

---

[88] J Hayes 'Is Clients' Use of AI Destroying Legal Privilege?' (Legal Futures, 16 July 2026), available at https://perma.cc/C8FZ-BX7H.

[89] BSB, *supra* n 43, section 6; for direct access clients, rC123.

[90] *Supra* n 4, para 16.

[91] The Law Society 'Generative AI – lifecycle considerations for SMEs' (2024), accompanying the guidance *supra* n 43.

## 6.5 Beyond England and Wales

The article is doctrinally English, but its structural claims attach to the technology rather than the law. Privilege and professional secrecy regimes protect what is confidential, and GenAI systems challenge confidentiality through the same three modes everywhere.

The extendibility is already visible. *Heppner* reached the *Munir* sentiment by orthodox American means (reasonable expectation of confidentiality defeated by the terms of use);[92] the American Bar Association's first ethics opinion on generative AI requires lawyers to evaluate the risk that client information will be disclosed to or accessed by others, requires that the client's informed consent be obtained before their confidential information is put into a tool that may use it for training, and warns that boilerplate consent in engagement letters will not do;[93] and the EDPB's position that trained models are not automatically anonymous makes the same point for EU data protection law.[94]

There are limits to that extendibility, of course. In EU competition investigations, for example, *Akzo Nobel* denies privilege to in-house lawyers' communications altogether, so in-house teams operating in that context must treat the AI analysis as an overlay on an already unprotected baseline.[95] And in civil law professional secrecy systems where the duty is owed by the lawyer and backed by criminal law rather than held by the client, the 'waiver' analysis transfers weakly. That said, the data-governance analysis transfers in full, since a secrecy obligation is breached by the leak itself, whatever the waiver doctrine. The portable core is this: wherever protection depends on demonstrable confidentiality, the three modes of data persistence determine what must be governed.

What does adjusted governance look like in concrete terms? For memorisation: know each tool's training policy and do not fine-tune on client data while selective erasure remains unsolved. For the context window: use enterprise-grade tools configured to match their terms; probe for context rot near the end of long documents; disable or scope prompt caching; govern shared workspaces so that membership tracks matter access rights; and treat externally sourced documents as prompt injection risks. For RAG: disable or partition semantic caching; set retention schedules for embeddings; ensure access controls follow documents into the vector store; build per-matter pipelines; and verify retrieval sources on outputs that matter. Across all three: train staff to recognise these failure modes and maintain an incident-response protocol that includes self-reporting.

A word, finally, on what this article does not cover. The three modes describe a single system processing a user's data. Agentic deployments (where the system is given connectors to email, document management and third-party tools, or control of the user's device) reinstate the classic exposure risks at every connector, and add one of their own: the hand-off of accumulated context window content between products, which the industry is now moving to standardise.[96] Standardisation makes the hand-off auditable, but each connector is a further boundary at which these questions must be asked again.

---

[92] *Supra* n 3.

[93] American Bar Association, Standing Committee on Ethics and Professional Responsibility, Formal Opinion 512 'Generative Artificial Intelligence Tools' (29 July 2024), discussing Model Rules 1.1, 1.4 and 1.6.

[94] *Supra* n 32.

[95] *Akzo Nobel Chemicals Ltd v European Commission* Case C-550/07 P [2010] ECR I-8301, [2011] 2 AC 338.

## 7 Conclusion

Three modes, three sets of risks, one principle: wherever the client's data goes, the lawyer's professional obligations go with it. An AI system, however impressive, does not know whether it is processing privileged material, and a responsible data governance framework needs to account for that.

Our conclusions can thus be stated as follows:

(a) the doctrinal tests for privilege have not changed, and are unlikely to; what has changed is the factual substrate to which they apply, and the knowledge needed to apply them;

(b) putting client material into a consumer-grade tool—one whose terms permit retention, training use and onward disclosure—breaches the duty of confidentiality and since disclosure on such terms is inconsistent with maintaining confidence as against the provider (not because the material is thereby published), it defeats the precondition of privilege or, where the client authorises the upload, waives it, although how far that loss extends as against an opponent who is a stranger to the provider's terms remains open;

(c) on the limited-disclosure authorities, a properly configured enterprise-grade deployment—no training use, zero or minimal retention, confidentiality obligations holding down the sub-processing chain, and documentation to prove it—should not in itself imperil privilege, and the contrary suggestion in the early case law rests on obiter guidance and imprecise terminology;

(d) the persistence layers that lawyers do not see—fine-tuned parameters, shared workspaces, vector stores, caches—are where the unintuitive breaches will happen, and they must be governed with the same rigour as the document management system; and

(e) because these failure modes are now documented and publicised by the regulator, the professional bodies and the senior judiciary, the standard of reasonable care has begun to absorb them: yesterday's reasonable precautions may be tomorrow's negligence.

The challenge is not the artificial intelligence but the data that moves through systems not designed for transparency, in ways that remain counter-intuitive until made visible. Lawyers who understand the difference between a model's parameters, its context window and its retrieval stores can manage these risks and maintain confidentiality, wherever they practise. Increasingly, technological competence is about knowing which aspects of a system's design affect confidentiality, and so what to ask, and of whom, before the client's material goes into an AI tool. Not every lawyer need answer those questions personally, but every lawyer should know that they arise.

---

[96] B Ambrogi 'DeepJudge Releases an Open Protocol for Passing Users – and Their Context – Between AI Products' (LawNext, 13 August 2026), available at https://perma.cc/SJS3-FB9C, describing the Agent Handoff Protocol, under which the objective, supporting materials, conversation history and analysis pass between products while 'tools, credentials, and execution privileges do not'.